\documentclass{ifacconf}

\usepackage{graphicx}      %
\usepackage{natbib}        %

\makeatletter
\let\old@ssect\@ssect %
\makeatother

\usepackage{amsmath,amssymb,amstext}

\usepackage{resizegather}

\usepackage{pgfplots} %
\pgfplotsset{compat=newest} 
\pgfplotsset{plot coordinates/math parser=false}

\usepackage{physics}
\usepackage{siunitx}

\usepackage{array}
\usepackage{algorithm}
\usepackage{etoolbox}

\let\classAND\AND
\let\AND\relax
\usepackage{algorithmic}

\let\AND\classAND
\AtBeginEnvironment{algorithmic}{\let\AND\algoAND}

\usepackage{xcolor}
\usepackage{mathtools}

\usepackage[hyperindex,breaklinks]{hyperref}
\hypersetup{
	colorlinks=true,
	linkcolor=blue,
	filecolor=magenta,      
	urlcolor=blue,
	citecolor=black,
}

\usepackage{enumitem}

\makeatletter
\def\@ssect#1#2#3#4#5#6{%
	\NR@gettitle{#6}%
	\old@ssect{#1}{#2}{#3}{#4}{#5}{#6}%
}
\makeatother

\usepackage{algorithm}
\usepackage{algorithmic}

\newcommand{\ivar}{{i_j}}

\newcommand{\change}[2]{{#2}}

\newcommand{\auxvar}{\tau_{\max}}

\newcommand{\lvar}{\Lambda}
\newcommand{\tpar}{p}
\newcommand{\refer}{\text{\normalfont ref}}

\newcommand{\itilde}{1}
\newcommand{\tnn}{0,0}

\newcommand{\svar}{t}
\newcommand{\tvar}{s}

\newcommand{\R}{\mathbb{R}}
\newcommand{\N}{\mathbb{N}}

\newcommand{\arxiv}[2]{#2}

\DeclareMathOperator\dom{dom}
\DeclareMathOperator\arctanh{arctanh}

\def\QEDclosed{\mbox{\rule[0pt]{1.3ex}{1.3ex}}} %

\def\qed{\QEDclosed} %

\makeatletter
\newcommand{\pushright}[1]{\ifmeasuring@#1\else\omit\hfill$\displaystyle#1$\fi\ignorespaces}
\newcommand{\pushleft}[1]{\ifmeasuring@#1\else\omit$\displaystyle#1$\hfill\fi\ignorespaces}
\makeatother

\newtheorem{defi}{Definition}
\newtheorem{theo}{Theorem}
\newtheorem{rema}{Remark}

\newtheorem{asum}{Assumption}
\begin{document}
	\begin{frontmatter}
		\title{	Self-triggered output feedback control for nonlinear networked control systems based on hybrid Lyapunov functions\thanksref{footnoteinfo}
		}
		
		\thanks[footnoteinfo]{		
			\change{
			Funded by Deutsche
			Forschungsgemeinschaft (DFG, German Research Foundation) under Germany’s
			Excellence Strategy - EXC 2075 - 390740016 and under grant
			AL 316/13-2 - 285825138. We acknowledge the support by the Stuttgart
			Center for Simulation Science (SimTech).}{$\copyright$ 2023 the authors. This work has been accepted to IFAC for publication under a Creative Commons Licence CC-BY-NC-ND.\\
			F. Allgöwer is thankful that this work was funded by the Deutsche Forschungsgemeinschaft (DFG, German Research Foundation)  under Germany’s Excellence Strategy -- EXC 2075 -- 390740016 and under grant AL 316/13-2 - 285825138.}
		}
		
		\author[first]{Michael Hertneck} 
		\author[first]{Frank Allg\"ower} 
		
		\address[first]{University of Stuttgart, Institute for Systems Theory and Automatic Control, Stuttgart, Germany (email: $\{$hertneck, allgower$\}$@ist.uni-stuttgart.de)}

		\begin{abstract}                %
			Most approaches for self-triggered control (STC) of nonlinear networked control systems (NCS) require measurements of the full system state to determine transmission times. However, for most control systems only a lower dimensional output is available. To bridge this gap, we present in this paper an output-feedback STC approach for nonlinear NCS. An asymptotically stable observer is used to reconstruct the plant state and transmission times are determined based on the observer state. 
			The approach employs hybrid Lyapunov functions and a dynamic variable to encode past state information and to maximize the time between transmissions. 
			 It is non-conservative in the sense that the assumptions on plant and controller are the same as for dynamic STC based on hybrid Lyapunov functions with full state measurements and any asymptotically stabilizing observer can be used.  We conclude that the proposed STC approach guarantees asymptotic stability of the origin for the closed-loop system.
		\end{abstract}
		
	
	\end{frontmatter}

\section{Introduction}
\label{sec_intro}
For networked control systems (NCS) with limited communication bandwidth, event-triggered control (ETC) and self-triggered control (STC) have emerged as key techniques to trade-off control performance and the usage of communication resources. In ETC, a state-dependent trigger rule is monitored continuously and a transmission 
 is triggered as soon as the trigger rule is violated. In STC, the controller determines at each transmission time based on available state information when the next transmission should take place.

In \cite{mazo2009self,anta2010sample}, it was demonstrated that the network load for NCS can be significantly reduced by STC compared to classical periodic sampling. For linear systems, a wide variety of STC strategies has been proposed, see, e.g., \cite{heemels2012introduction} and the references therein. For nonlinear systems, a smaller but increasing number of approaches is available. In \cite{benedetto2013digital,tiberi2013simple,theodosis2018self}, Lipschitz continuity properties are exploited to determine transmission times such that a decrease of a Lyapunov function can be guaranteed. Small gain techniques are used in \cite{tolic2012self,liu2015small}.  In \cite{anta2010sample,delimpaltadakis2020isochronous,delimpaltadakis2020region}, isochroneity properties of homogeneous systems are leveraged. In \cite{hertneck21robust_arxiv}, hybrid Lyapunov functions and a dynamic variable that encodes the past system behavior are used to determine transmission times.

Whilst there are several approaches for the design of STC mechanisms for linear systems with output feedback regulators, see, e.g., \cite{almeida2014self,gleizer2020self}, the aforementioned STC approaches for nonlinear systems all require information about the full plant state and are thus limited to state feedback. However, in most practical scenarios only a lower dimensional output can be used for control. 

To bridge this gap, we present in this paper an output feedback STC approach for nonlinear NCS. The approach is deduced from the dynamic STC approach based on hybrid Lyapunov functions with state-feedback from \cite{hertneck21robust_arxiv}. A continuous observer that is located at the sensor node is employed to reconstruct the plant state based on the plant output. The observer can, e.g., be designed using one of the methods from \cite{bernard2022observer}. Transmission times and plant inputs are determined based on the observer state. The proposed approach is non-conservative in the sense that the assumptions on plant and controller are the same as in the state-feedback case from \cite{hertneck21robust_arxiv} and any nonlinear observer that ensures asymptotic stability of the origin for the observer error can be used. We present the modifications that are required for the proposed dynamic output feedback STC approach and model the overall NCS as a hybrid system. For this system, we prove asymptotic stability of the origin. We illustrate the proposed approach with a numerical example.

The remainder of this paper is structured as follows.  In Section~\ref{sec_setup}, we present the setup of the paper and specify our control objective. Some preliminaries are discussed in Section~\ref{sec_prelim}. In Section~\ref{sec_main}, we detail the proposed STC approach and derive stability guarantees. A numerical example is given in Section~\ref{sec_ex}. Section~\ref{sec_conc} concludes the paper. 

\subsection*{Notation and definitions}
The nonnegative real numbers are denoted by  $\mathbb{R}_{\geq 0} $. The natural numbers are denoted by $\mathbb{N}$, and we define $\mathbb{N}_0:=\mathbb{N}\cup  \left\lbrace 0 \right\rbrace $. 
We denote the Euclidean norm by $\abs{\cdot}$.
 A continuous function $\alpha: \mathbb{R}_{\geq 0} \rightarrow \mathbb{R}_{\geq 0}$ is a class $ \mathcal{K}$ function if it is strictly increasing and $\alpha(0) = 0$. It is a class $\mathcal{K}_\infty$ function if it is a class $\mathcal{K}$ function and it is unbounded. A continuous function $\beta:\mathbb{R}_{\geq 0}\times \mathbb{R}_{\geq 0} \rightarrow \mathbb{R}_{\geq 0}$ is a class $\mathcal{K}\mathcal{L}$ function, if $\beta(\cdot,t)$ is a class $\mathcal{K}$ function for each $t \in\mathbb{R}_{\geq 0}$ and $\beta(q,\cdot)$ is nonincreasing and satisfies $\lim\limits_{t \rightarrow \infty} \beta(q,t) = 0$ for each $q\in\mathbb{R}_{\geq 0}$. A function $\beta:\mathbb{R}_{\geq 0}\times \mathbb{R}_{\geq 0} \times \mathbb{R}_{\geq 0} \rightarrow \mathbb{R}_{\geq 0}$ is a class $\mathcal{K}\mathcal{L}\mathcal{L}$ function if for each $r \geq 0$, $\beta(\cdot,r,\cdot)$ and $\beta(\cdot,\cdot,r)$ are class $\mathcal{K}\mathcal{L}$ functions.

We use \cite[Definitions 1-3]{carnevale2007lyapunov}, that are originally taken from \cite{goebel2006solutions}, to characterize a hybrid model of the considered NCS and corresponding hybrid time domains, trajectories and solutions. Moreover, we adapt the definitions of maximal solutions
 and $t-$completeness from \cite{goebel2006solutions}.
\section{Setup}
\label{sec_setup}
We consider a setup where the sensors and actuators of a continuous plant are connected through a communication network and exchange information only at discrete transmission times.
The plant is described by
\begin{equation}
	\label{eq_plant}
	\begin{split}
		\dot{x}_p &= f_p(x_p,\hat{u}),\\
		y &= g_p(x_p)
	\end{split}	
\end{equation}
where $x_p(t) \in \R^{n_x}$ is the plant state with initial condition  $x_p(0)$, $y(t)\in\R^{n_y}$ is the plant output and $\hat{u}(t) \in \R^{n_u}$ is the last input that has been received by the plant.

Only the output $y$ but not the whole plant state $x_p$ can be measured. Thus, an observer that is connected to the plant's sensors is used to reconstruct the plant state based on $y$. The observer is described\footnote{The observer in \eqref{eq_obs} uses continuous feedback. Note that for suitable system classes, a sampled-data observer as, e.g., in \cite{raff2008observer} could be used as well.} by 
\begin{equation}
	\label{eq_obs}
	\dot{x}_o = f_o(x_o,\hat{u},y),
\end{equation}
where $x_o(t)\in\R^{n_x}$ is the observer state.  

 The input is generated by the static state-feedback controller
 \begin{equation*}
 	u = g_c(x_o)
 \end{equation*}
 using the observer state $x_o$.
In this paper, we consider an emulation approach, i.e., we suppose that the controller has been designed for state feedback for plant~\eqref{eq_plant} ignoring the network effects using any method for the design of nonlinear continuous state-feedback controllers with some robustness to the input errors.
The functions $f_p$ and $f_o$ are assumed to be continuous and the functions $g_p$ and $g_c$ are assumed to be continuously differentiable.

The transmission times $(\svar_j)_{j\in\mathbb{N}_0}$ are determined by an STC mechanism to be specified later. 
At each transmission time, the input $\hat{u}$ is updated based on the current values of $g_c(x_o),$ i.e., $\hat{u}(\svar_j) = u(\svar_j) = g_c(x_o(t_j))$. Further, we denote by $\hat{x}_o$ the observer state associated to the last transmission time, i.e., $\hat{x}_o(\svar_j) = x_o(\svar_j)$.   Between transmission times, we assume that $\hat{u}$ (and thus $\hat{x}_o$) is kept constant, which resembles a zero-order-hold (ZOH) scenario. 
We introduce $e\coloneqq \hat{x}_o - x_p$ with $e(t) \in \R^{n_x}$ to describe  the sum of the observer induced error and the network induced error.

Similar as in \cite{hertneck21robust_arxiv}, we consider in this paper a dynamic STC mechanism that determines at transmission times $\svar_j$ the next transmission time $\svar_{j+1}$ based on information that is available to the STC mechanism at $t_j$ including an internal state $\eta$. The internal state  $\eta(t)\in\R^{n_\eta}$ is used to incorporate the past system behavior when deciding about the next transmission time. Since we consider an output feedback scenario, 
the plant state cannot be directly used to determine transmission times.
Instead, the proposed STC mechanism uses the observer state $x_o$ to determine transmission times.
It can thus be described by 
$\svar_{j+1} \coloneqq \svar_j + \Gamma(x_o(\svar_j),\eta(\svar_j)),$
where $\Gamma:\R^{n_x}\times\R^{n_\eta} \rightarrow \left[\svar_{\min},\infty\right)$ for some $\svar_{\min} > 0$. We will later provide an explicit value for $t_{\min}$ for the mechanism. 

The dynamic variable $\eta$ is updated at  transmission times based on its current value and the current observer output, and remains constant in between  transmission times. Thus, $\eta$ evolves according to 
\begin{equation}
	\begin{split}
		\eta(\svar_{j+1}) = S(\eta(\svar_j),x_o(\svar_j)),
		\\
		\dot{\eta}(t) = 0, t\in\left[\svar_j,\svar_{j+1}\right)
	\end{split}
\end{equation}
for some $\eta(0)$, where $S:\R^{n_\eta}\times\R^{n_x} \rightarrow \R^{n_\eta}.$

In order to model the overall networked control systems as a hybrid system, we introduce the timer variable $\tau$ which keeps track of the elapsed time since the last  transmission time and the auxiliary variable $\auxvar$ which encodes the next  transmission time. Using this, we obtain
\begin{equation}
\label{eq_sys_hyb}
	\begin{cases}
		\dot{\xi} = F(\xi), & \xi \in C,\\
		\xi^+ = G(\xi), & \xi \in D,
	\end{cases}
\end{equation}
with $\xi \coloneqq \left[x_p^\top,x_o^\top,e^\top,\eta^\top,\tau,\auxvar\right]^\top,$ 
	$F(\xi) \coloneqq \linebreak\left(f(x_p,e)^\top,f_o(x_o,g_c(x_p+e),g_p(x_p)),g(x,e)^\top,0,1,0\right)^\top,$
\linebreak where
$	f(x_p,e) = f_p(x_p,g_c(x_p+e)) $
and $g(x,e) = -f(x,e)$,
	$G(\xi) \coloneqq \left(x_p^\top,x_o^\top,0,S(\eta,x)^\top,0,\Gamma(x,\eta)\right)^\top,$
and with \linebreak
$	C := \left\lbrace \xi \in \R^{2n_x+n_e+n_\eta+2} | \tau \leq \auxvar \right\rbrace$ and $
	D := \linebreak \left\lbrace \xi \in \R^{2n_x+n_e+n_\eta+2} | \tau = \auxvar \right\rbrace.$

Jumps of the hybrid system~\eqref{eq_sys_hyb} correspond for any solution~$\xi$ exactly to transmission times of the STC mechanism. Hence the transmission sequence $(t_j,j)\in\dom~\xi$ corresponds exactly to the indices when \eqref{eq_sys_hyb} jumps.  We thus describe by $\tvar_j \coloneqq  (\svar_j,j-1)$ the hybrid time before the transmission at time $\svar_j$ and by $\tvar_j^+ \coloneqq (\svar_j,j)$ the hybrid time directly after the transmission at time $\svar_j$. 
We assume that the STC mechanism is executed at the initial time $t_0 = 0$. This corresponds to a restriction of the initial conditions for the hybrid system for $e(0,0), \tau(0,0)$ and $\auxvar(0,0)$ to $e(0,0) = x_o(0,0)-x_p(0,0), \tau(0,0) = 0$ and $\auxvar(0,0) = \Gamma(x_o(0,0),\eta(0,0))$. Otherwise the first transmission time might not be well-defined.

Subsequently, our goal will be to design functions $\Gamma$ and $S$ that ensure input-to-state stability of the origin of \eqref{eq_sys_hyb} according to the following definition. 

\begin{defi}
	\label{def_asym_stab}
	For the hybrid system~$\mathcal{H}_{STC}$ with initial condition $x_p(0,0) \in \R^{n_x} , x_o(0,0)\in\R^{n_x}, e(0,0)  = x_o(0,0)-x_p(0,0), \eta(0,0) \in \R^{n_\eta}$ and $\auxvar(0,0) \linebreak = \Gamma(x_o(0,0),\eta(0,0))$, the set
	\begin{equation*}
		\left\lbrace\left(x_p,x_o,e,\eta,\tau,\auxvar\right):x_p = 0, x_o = 0, e= 0, \eta = 0 \right\rbrace
	\end{equation*} is uniformly globally asymptotically stable (UGAS), if there exists $\beta \in \mathcal{K}\mathcal{L}\mathcal{L}$ such that all corresponding maximal solutions $\xi$ are $t-$complete and satisfy for all $(t,j)\in\dom~\xi$
	\begin{equation*}
		\abs{\begin{bmatrix}
				x_p(t,j)\\
				e(t,j)\\
				\eta(t,j)
		\end{bmatrix}} \leq \beta\left(\abs{\begin{bmatrix}
				x_p(0,0)\\
				e(0,0)\\
				\eta(0,0)
		\end{bmatrix}},t, j\right).
	\end{equation*}
\end{defi} 
\section{Preliminaries}
\label{sec_prelim}
In this section, we present the assumptions that we make on plant, controller and observer and recap some preliminaries that are needed for the proposed STC approach. 

\subsection{Hybrid Lyapunov functions}

In this subsection, we recap assumptions on the plant and the controller that are needed to construct a hybrid Lyapunov function and to derive a bound on it. We use the following assumption \cite[Assumption~1 for $w = 0$]{hertneck21robust_arxiv}, that is based on \cite[Assumption~1]{carnevale2007lyapunov}.
\begin{asum}
	\label{asum_hybrid_lyap}
	There exist a locally Lipschitz function $W:\mathbb{R}^{n_e} \rightarrow \mathbb{R}_{\geq0}$, a locally Lipschitz function $V:\mathbb{R}^{n_x} \rightarrow \mathbb{R}_{\geq0}$, a continuous function $H:\mathbb{R}^{n_x}\times\mathbb{R}^{n_e} \rightarrow \mathbb{R}_{\geq0}$, constants $L, \gamma\in \mathbb{R}_{>0}$, $\epsilon\in \mathbb{R}$, and  $\underline{\alpha}_W$, $\overline{\alpha}_W, \underline{\alpha}_V, \overline{\alpha}_V\in \mathcal{K}_\infty$  such that for all $e\in\mathbb{R}^{n_e}$,
	\begin{equation}
		\label{eq_w_bound}
		\underline{\alpha}_W(\abs{e}) \leq W(e) \leq \overline{\alpha}_W(\abs{e}),
	\end{equation}
	for all $x_p \in\mathbb{R}^{n_x}$,
	\begin{equation}
		\label{eq_V_bound_K}
		\underline{\alpha}_V(\abs{x_p}) \leq V(x_p) \leq \overline{\alpha}_V(\abs{x_p}),
	\end{equation}
	and for all  $x_p \in \mathbb{R}^{n_x} $ and almost all $e\in\mathbb{R}^{n_e},$ 
	\begin{equation}
		\left\langle \frac{\partial W(e)}{\partial e},g(x_p,e)\right\rangle \leq L W(e) + H(x_p,e).  \label{eq_w_est}
	\end{equation}
	Moreover, for all $e \in \mathbb{R}^{n_e}$ and almost all $x_p \in \mathbb{R}^{n_x}$,  
	\begin{equation}
		\begin{split}
			&\left\langle \nabla V(x_p),f(x_p,e) \right\rangle\\
			\leq &- \epsilon V(x_p) -H^2(x_p,e)
			+ \gamma^2 W^2(e).
		\end{split}
		\label{eq_v_desc_hybrid}
	\end{equation}
\end{asum}

A discussion of this assumption can be found in \cite{carnevale2007lyapunov}. Note that it involves only the plant and the controller and does not depend on the observer. We will specify conditions on the observer in the next subsection. 
Note also that Assumption~\ref{asum_hybrid_lyap} can hold simultaneously for different choices of $\epsilon, \gamma$ and $L$. If we can find one set of parameters for which the assumption holds, then we will typically also be able to find many different parameter sets.

To determine transmission times, the proposed STC framework will use a bound on the evolution of $V(x)$, that is adapted from \cite{hertneck21robust_arxiv}. This bound is based on the function
\begin{equation}
	T_{\max}(\gamma,\lvar ) \coloneqq \begin{cases}\vspace{1mm}
		\frac{1}{\lvar r} \arctan(r) & \gamma > \lvar \\ \vspace{1mm}
		\frac{1}{\lvar } & \gamma = \lvar \\
		\frac{1}{\lvar r} \arctanh(r) &\gamma < \lvar 
	\end{cases}
\end{equation}
where
	$r\coloneqq\sqrt{\abs{
			\left(\frac{\gamma}{\lvar }\right)^2-1}},$
that was originally used in \cite{nesic2009explicit} to determine the maximum allowable sampling interval for sampled-data systems. We use the following result from \cite{hertneck21robust_arxiv}.

\begin{prop}{\cite[Proposition~1 for $w = 0$]{hertneck21robust_arxiv}}.
	\label{prop_hybrid}
	Consider any maximal solution $\xi$ to \eqref{eq_sys_hyb} at transmission time $\tvar_j^+$ for $j \in \change{\mathbb{N}_0}{\dom~\xi}$.  Let Assumption~\ref{asum_hybrid_lyap} hold 
	for some $\gamma, \epsilon$ and $L$.
	Moreover, let $0 < \auxvar(\tvar_j^+) < T_{\max} (\gamma,\max\left\lbrace L+\frac{\epsilon}{2},1-\delta\right\rbrace )$  for $\delta\in\left(0,1\right)$.	
	Consider 
	\begin{equation}
		\label{eq_def_u}
		U(\xi) \coloneqq V(x)+\gamma \phi(\tau) W^2(e),
	\end{equation}
	where 	 $\phi : [0,\auxvar(\tvar_j^+)] \rightarrow \mathbb{R}$ is the solution to
	\begin{equation}
		\label{eq_def_phi}
		\dot{\phi} = -2\max\left\lbrace L+\frac{\epsilon}{2},1-\delta\right\rbrace \phi-\gamma(\phi^2+1),~ \phi(0) = \lambda^{-1}
	\end{equation}
	for some sufficiently small $\lambda \in \left(0,1\right)$.
	Then, \change{}{$t_{j+1} =  \svar_j+\auxvar(\tvar_j^+) \in \dom~\xi$} and for all $\svar_j \leq t \leq \change{\svar_j+\auxvar(\tvar_j^+)}{t_{j+1}}$, it holds that	
	\begin{equation}
		\begin{split}
			V(x_p(\svar,j\change{+1}{}))
			\leq& U(\xi(\svar,j\change{+1}{}))
			\leq	e^{ -\epsilon (t-\svar_j)}U(\xi(\tvar_j^+)).\\
		\end{split}
		\label{eq_prop_hybrid1}
	\end{equation}
\end{prop}	
		
Proposition~\ref{prop_hybrid} yields an upper bound on the evolution of $U(\xi)$ for the parameters $\epsilon,\gamma$ and  $L$ that satisfy Assumption~\ref{asum_hybrid_lyap}. Thus, it also provides an upper bound on $V(x)$ due to $U(\xi) \geq V(x)$.
  This bound is valid, if the time between two transmissions is bounded by $T_{\max}(\gamma,\max\left\lbrace L-\frac{\epsilon}{2},1-\delta\right\rbrace)$. The actual bound depends on the parameters from Assumption~\ref{asum_hybrid_lyap}. Particularly, if $\epsilon > 0$, then the bound is exponentially decreasing over time. In contrast, if $\epsilon < 0$, then the bound is increasing. However, the admissible time between transmissions $T_{\max}(\gamma,L+\frac{\epsilon}{2})$ decreases when $\gamma$ or $\epsilon$ are increased.  
We thus observe in Proposition~\ref{prop_hybrid} a trade-off between the admissible time between transmissions and the growth of the bound on $V(x)$. Particularly, if the time between two successive transmissions is small, then we will be able to choose $\epsilon$ 
large and thus obtain an exponentially decreasing bound on $V(x)$. In contrast, if the time between two successive transmissions is large, then we need to choose
 $\epsilon$ small to be able to derive a bound on $V(x)$, which has the effect that this bound may be increasing. 

Note that $U(\xi(\tvar_j^+))$ depends on $e(\tvar_j^+)$ and thus on the observer error (whilst the network induced error has been reset to $0$ at time $s_j^+$). However, the observer error is not available to the STC mechanism and can thus not  be used to determine the next transmission time. Instead, the STC mechanism will use the value of $V(x_o)$ to determine transmission times. To obtain still stability guarantees, we will need an assumption on the observer which we introduce in the next subsection.

\subsection{Bound on the observer error}
We assume that the observer is designed such, that the observer error $e_o = x_o-x_p$ is asymptotically stable. More formally, we make the following assumption.
\begin{asum}
	\label{as_obs_er}
	The observer~\eqref{eq_obs} is such that 
	\begin{equation}
		\label{eq_obs_er}
		\abs{x_o(t,j)-x_p(t,j)} = \abs{e_o(t,j)}  \leq \beta_o(\abs{e_o(0,0)},t)
	\end{equation}
	holds for $\beta_o\in\mathcal{K}\mathcal{L}$. 
\end{asum}

Assumption~\ref{algo_trig_window} requires the observer to be exponentially stable. Note that the design of observers for nonlinear continuous-time systems is not trivial but widely studied in the literature (see, e.g., \cite{bernard2022observer}) and thus beyond the scope of this paper. 
\section{Dynamic output feedback STC}
\label{sec_main}
In this section, we present the details of the dynamic output feedback STC approach and give stability guarantees. The proposed approach is a modified version of the dynamic STC approach that requires full state measurements from \cite{hertneck21robust_arxiv}. We will first briefly recap the main idea of the dynamic STC mechanism with full state measurements from \cite{hertneck21robust_arxiv} and then explain how it can be modified to still work with output feedback.
We assume throughout this section that $\xi$ is a maximal solution to \eqref{eq_sys_hyb}. 

\subsection{Recap: Dynamic STC with full state measurements}
In the case with full state measurement from \cite{hertneck21robust_arxiv}, the strategy is to choose, at a transmission time $\tvar_j$, the transmission time $\tvar_{j+1}$ such that $V(x_p(\tvar_{j+1}))$ does not exceed a discounted average of its past $m$ values for some chosen $m\in\N$, i.e., the ideal goal is to choose $s_{j+1}$
 such that 
\begin{equation}
	\label{eq_idea_window}
	V(x_p(\tvar_{j+1})) \leq \frac{1}{m} \sum_{k=j-m+1}^{j} e^{-\epsilon_\refer \left(t_{j+1} - t_k\right)} V(x_p(\tvar_k))
\end{equation}
hols for some $\epsilon_\refer > 0$.
This choice can be implemented in our hybrid system model \eqref{eq_sys_hyb} by choosing $n_\eta = m-1$ as the dimension of the dynamic variable and define the update rule of the dynamic variable as\footnote{Note that we use here abusively $S(\eta,x_p)$ and $\Gamma(x_p,\eta)$ instead of $S(\eta,x_o)$ and $\Gamma(x_o,\eta)$ since we recap dynamic STC with full state measurements in this subsection.}
\begin{equation}
	\label{eq_S_window}
	S(\eta,x_p) = \begin{pmatrix}
		e^{-\epsilon_\refer\Gamma(x_p,\eta)}\eta_2\\
		\vdots\\
		e^{-\epsilon_\refer\Gamma(x_p,\eta)}\eta_{m-1}\\
		e^{-\epsilon_\refer\Gamma(x_p,\eta)}V(x_p)
	\end{pmatrix}.
\end{equation}
 Note that  $t_{j+1}-t_{j} = \auxvar(\tvar_j^+) = \Gamma(x_p(\tvar_j),\eta(\tvar_j))$. Hence, $\eta_k(s_j) = e^{-\epsilon_\refer (t_j-t_{j-m+k})} V(x_p(s_{j-m+k}))	$ holds for this choice of $S(\eta,x)$,
 if $k > m-j$, which implies
$V(x_p(\tvar_j))+\sum_{k=1}^{m-1} \eta_k(\tvar_j) = \sum_{k = j-m+1}^{j} e^{\epsilon_\refer \left(t_j - t_k\right)} V(x_p(s_k))$
for $j > m$.  

Due to Proposition~\ref{prop_hybrid}, \eqref{eq_idea_window} is satisfied, if there is one parameter set $(\epsilon_i,\gamma_i,L_i)$ satisfying Assumption~\ref{asum_hybrid_lyap} for which 
$e^{ -\epsilon_i\auxvar(\tvar_j^+)} V(x_p(\tvar_j))
\leq e^{-\epsilon_\refer\auxvar(\tvar_j^+)} C(x_p(\tvar_j),\eta(\tvar_j)) $
and $\auxvar(\tvar_j^+) < T_{\max}(\gamma_i,\max\left\lbrace L_i+\frac{\epsilon_i}{2},1-\delta\right\rbrace)$ hold, where $	C(x_p,\eta) = V(x_p) + \sum_{k=1}^{m-1} \eta_k.$
Given $(\epsilon_i,\gamma_i,L_i)$, it can thus be ensured that \eqref{eq_idea_window} holds if the next transmission time is selected as
\begin{equation}
	\begin{split}
		\auxvar(\tvar_j^+)  =& \min\left\lbrace \delta T_{\max}\left(\gamma_i,\max\left\lbrace L_i+\frac{\epsilon_i}{2},1-\delta\right\rbrace\right)\vphantom{\frac{\log(C(x_p(\tvar_j),\eta(\tvar_j)))-\log(V(x_p(\tvar_j)))}{\max\left\lbrace -\epsilon_i,2(L_i-\lvar_i ) \right\rbrace + \epsilon_\refer}}\right.,\\
		&\left.\frac{\log(C(x_p(\tvar_j),\eta(\tvar_j)))-\log(V(x_p(\tvar_j)))}{ -\epsilon_i + \epsilon_\refer} \right\rbrace,
			\label{eq_trigger_t1}
	\end{split}
\end{equation}
see \cite[Section III.B]{hertneck21robust_arxiv} for a detailed derivation. The idea for the dynamic STC mechanism is now to compute offline $n_\tpar$ different parameter sets $(\epsilon_i,\gamma_i,L_i), i\in\left\lbrace 1,\dots,n_\tpar\right\rbrace$ for which Assumption~\ref{asum_hybrid_lyap} holds and to maximize $\auxvar(\tvar_j^+)$  in \eqref{eq_trigger_t1} over all parameter sets to determine the next transmission time. If this is not possible, then a fallback strategy is used, that is to choose the next transmission time sufficiently small to still guarantee a decrease of $V$, which is always possible if Assumption~\ref{asum_hybrid_lyap} holds with some $\epsilon_i > 0$. 
The complete procedure to select the next transmission time in the state-feedback case is summarized in \cite[Algorithm~2]{hertneck21robust_arxiv}.
It has been shown in \cite{hertneck21robust_arxiv} that the resulting dynamic STC mechanism guarantees UGAS and that it can significantly reduce the number of transmissions in comparison to periodic time-triggered sampling. 
However $V(x_p(\tvar_j))$ and its past $m-1$ values are needed to compute $\auxvar(\tvar_j^+)$ in \eqref{eq_trigger_t1}, which requires full state feedback and is not possible if only a lower dimensional output is available. We describe in the next subsection, how the observed state $x_o(\tvar_j)$ can be used instead to determine transmission times. 

\subsection{Modifications for dynamic output feedback STC and stability result}
Since neither $V(x_p(\tvar_j))$ nor its values from the past $m-1$ transmission times can be measured, we use instead $V(x_o(\tvar_j))$ and the values of the Lyapunov function from the past $m-1$ transmission times at transmission time $\tvar_j$ to determine the next transmission time $\tvar_{j+1}$. To do so, the following changes are needed in comparison to the state-feedback variant of the STC mechanism. 
First, we change the update of the dynamic variable to 
\begin{equation}
	\label{eq_S_window_2}
	S(\eta,x_o) = \begin{pmatrix}
		e^{-\epsilon_\refer\Gamma(x_o,\eta)}\eta_2\\
		\vdots\\
		e^{-\epsilon_\refer\Gamma(x_o,\eta)}\eta_{m-1}\\
		e^{-\epsilon_\refer\Gamma(x_o,\eta)}\min\left\lbrace V(x_o), V_{\max}\right\rbrace 
	\end{pmatrix},
\end{equation}
i.e., we use $V(x_o)$ instead of $V(x_p)$ and we bound each component of the dynamic variable by some chosen  $V_{\max}\in\R_{>0}$. The latter is needed in case that the plant state grows faster than the observer state decreases to limit the maximum increase of the plant state.
Second, we modify the procedure to determine transmission times. We replace \eqref{eq_trigger_t1} by
\begin{equation}
	\begin{split}
		\auxvar(\tvar_j^+)  =& \min\left\lbrace \delta T_{\max}\left(\gamma_i,\max\left\lbrace L_i+\frac{\epsilon_i}{2},1-\delta\right\rbrace\right)\vphantom{\frac{\log(C(x_o(\tvar_j),\eta(\tvar_j)))-\log(V(x_o(\tvar_j)))}{\max\left\lbrace -\epsilon_i,2(L_i-\lvar_i ) \right\rbrace + \epsilon_\refer}}\right.,\\
		&\left.\frac{\log(C(x_o(\tvar_j),\eta(\tvar_j)))-\log(V(x_o(\tvar_j)))}{ -\epsilon_i + \epsilon_\refer} \right\rbrace,
		\label{eq_trigger_t2}
	\end{split}
\end{equation} 
where we use 
\begin{equation}
	\label{eq_C_def}
	C(x_o,\eta) = \frac{1}{m}\left( V(x_o) + \sum_{k=1}^{m-1} \eta_k\right).
\end{equation}
	\begin{algorithm}[tb]
		\caption{Computation of $\Gamma(x_o,\eta)$  for some $\delta \in \left(0,1\right)$, some $V_{\max}\in\R_{>0}$ and given $C(x_o,\eta)$. }
		\label{algo_trig_window}
		\begin{algorithmic}[1]
			\STATE $V \leftarrow V(x_o)$, $C \leftarrow C(x_o,\eta)$ %
			\STATE $\bar h \leftarrow \delta T_{\max}\left(\gamma_\itilde,L_\itilde+\frac{\epsilon_\itilde}{2}\right)$ \label{line_fallback}
			\FOR{\text{\bf each} $i \in \left\lbrace2,\dots,n_\tpar\right\rbrace$ } 
			\STATE $\Lambda_i \leftarrow \max \left\lbrace L_i + \frac{\epsilon_i}{2},(1-\delta) \right\rbrace$
			\IF{$V_{\max} \geq C \geq V$} \label{line_for_start} %
			\IF{$-\epsilon_i+\epsilon_\refer > 0$}
			\STATE $\bar h_i \leftarrow \min\left\lbrace \delta T_{\max}(\gamma_i,\Lambda_i),				\frac{\log(C)-\log(V)}{ -\epsilon_i+ \epsilon_\refer} \right\rbrace$ \label{line_hi}
			\ELSE
			\STATE $\bar h_i \leftarrow \delta T_{\max}(\gamma_i,\Lambda_i)$
			\ENDIF
			\ELSE
			\STATE $\bar h_i \leftarrow 0$
			\ENDIF\label{line_for_end}
			\IF{$\bar h_i > \bar h$}
			\STATE $\bar h \leftarrow \bar h_i$\label{line_h_update}
			\ENDIF
			\ENDFOR 			
			\STATE $\Gamma(x_o,\eta) \leftarrow \bar{h}$
		\end{algorithmic}
	\end{algorithm}
The modified procedure to select transmission times is described by Algorithm~\ref{algo_trig_window}, which is a modified variant of \cite[Algorithm~2]{hertneck21robust_arxiv}. Note that an additional important modification in the algorithm, that is needed for technical reasons to obtain stability guarantees, is that the algorithm selects $\bar{h} > \delta T_{\max}\left(\gamma_\itilde,L_\itilde+\frac{\epsilon_\itilde}{2}\right)$ only if $C(x_o(\tvar_j),\eta(\tvar_j)) \leq V_{\max}$ for the chosen $V_{\max}$. This means that the algorithm resorts to the fallback strategy if the value of the observed Lyapunov function exceeds the bound $V_{\max}$. It is needed if $x_p$ and $x_o$ grow faster than $e_o$ converges to limit the maximum increase of $x_p$ and $x_o$. 
We obtain the following result for the proposed dynamic STC mechanism with output feedback. 
\begin{theo}
	\label{prop_fir}
	Consider the hybrid system \eqref{eq_sys_hyb} with $S(\eta,x_o)$ defined according to \eqref{eq_S_window_2} and $\Gamma(x_o,\eta)$ defined by Algorithm~\ref{algo_trig_window} with $C(x_o,\eta)$ according to \eqref{eq_C_def}, some $\delta \in 
		\left(0,1\right)$ and $V_{\max} >0$. Assume there are $n_\tpar$ different parameter sets $\epsilon_i, \gamma_i, L_i$, $i \in \left\lbrace 1,\dots,n_\tpar\right\rbrace$, for which Assumption~1 holds with the same function $V$ and let $\epsilon_1\geq\epsilon_\refer > 0$. Let Assumption~\ref{as_obs_er} hold. Then, the set $\left\lbrace \left(x_p,x_o,e,\eta,\tau,\auxvar\right): x_p = 0, x_o = 0, e= 0, \eta = 0 \right\rbrace$ is UGAS and for any maximal solution $\xi$, $\svar_{j+1}- \svar_j \geq t_{\min} \coloneqq \delta T_{\max}\left(\gamma_1,L_1+\frac{\epsilon_1}{2}\right)$ for all $j\in\change{\mathbb{N}_0}{\dom~\xi}$.
\end{theo}
\arxiv{The proof ot Theorem~\ref{prop_fir} can be found in the preprint \url{https://arxiv.org/abs/xxxx.xxxxx}.}{The proof of Theorem~\ref{prop_fir} is given in Appendix~\ref{app_a}.}

\begin{rema}
	The main difference of the proposed output feedback STC mechanism in comparison to the variant with full state measurements is that the observer state is used instead of the real plant state to determine transmission times and to compute inputs that are applied to the plant. Since the observer error is handled together with the network induced error, Assumption~\ref{asum_hybrid_lyap} is the same as for the full state-feedback case (cf. \cite[Assumption~1 for $w=0$]{hertneck21robust_arxiv}), i.e., the parameters $\epsilon_i,\gamma_i,L_i$ can be determined in the same way as for the case with full state measurements and are independent from the observer choice. Since the observer has no influence on Assumption~\ref{asum_hybrid_lyap}, the design of controller and STC mechanism can be separated from the design of the observer, and any observer that satisfies Assumption~\ref{as_obs_er} can be used.
\end{rema}

\begin{rema}
	There are three parameters $\epsilon_\refer, V_{\max}$ and $\delta$ for the proposed STC mechanism that can be chosen by the user. The worst case asymptotic convergence speed of the system state is determined by $\epsilon_\refer$. The bound $V_{\max}$ is needed to avoid that the Lyapunov function grows faster than the observer error converges and can be chosen arbitrarily large. The parameter $\delta$ is needed for technical reasons. For a preferably large time between transmissions, it should be chosen close to $1$, e.g., $\delta = 0.999$. 
\end{rema}

\begin{rema}
	Note that the proof of Theorem~\ref{prop_fir} can be modified if $\epsilon V(x)$ is replaced in Assumption~\ref{asum_hybrid_lyap} for $i = 1$ by $\rho(\abs{x})$ for some $\rho\in\mathcal{K}$ to still guarantee UGAS. This makes the assumption less restrictive but requires however some additional technical modifications that are omitted for simplicity. 
\end{rema}

\section{Example}
\label{sec_ex}
In this section, we illustrate the proposed STC approach with a numerical example from the literature. We consider the single link robot arm example from \cite{postoyan2014tracking}. The plant is given by 
\begin{equation*}
	\begin{split}
		x_{p,1}&= x_{p,2}\\
		x_{p,2}&= -a\sin(x_{p,1}) + b\hat{u}\\
		y&=x_{p,1}
	\end{split}	
\end{equation*}
and the static state-feedback controller by \linebreak
	$u = b^{-1}\left(a\sin(x_{o,1})-x_{o,1}-x_{o,2}\right).$
We consider an observer given by 
\begin{equation*}
	\begin{split}
		x_{o,1} &= x_{o,2} + \theta_1\left(y-x_{o,1}\right)\\
		x_{o,2} &= -a\sin(x_{o,1}) +b\hat{u} + \theta_2\left(y-x_{o,1}\right)
	\end{split}
\end{equation*}
for $\theta_1,\theta_2\in\R$. We define $x_p = \begin{bmatrix}
	x_{p,1} & x_{p,2}
\end{bmatrix}^\top$, $e = \begin{bmatrix}
	e_1 & e_2
\end{bmatrix}^\top = \begin{bmatrix}
	\hat{x}_{o,1} - x_{p,1} & \hat{x}_{o,2} - x_{p,2}
\end{bmatrix}^\top$
and $e_o = \begin{bmatrix}
	e_{o,1} & e_{o,2}
\end{bmatrix}^\top = \linebreak \begin{bmatrix}
	{x}_{o,1} - x_{p,1} & {x}_{o,2} - x_{p,2}
\end{bmatrix}^\top$.

Note again that Assumption~\ref{asum_hybrid_lyap} is identical as for the state-feedback case and can thus be verified as in state-feedback case based on linear matrix inequalities. Details on the verification procedure for the considered example are omitted for brevity but can be found in \cite[Section VI.A]{hertneck21robust_arxiv}.
Using 
\begin{equation*}
	\begin{split}
		&-a\left(\sin(x_{p,1}) - \sin(x_{o,1})\right)\\
		=&2a\cos\left(\frac{2x_{p,1}+e_{o,1}}{2}\right) \sin\left(-\frac{e_{o,1}}{2}\right) = \tilde{a} e_{o,1}
	\end{split}
\end{equation*}
for a varying parameter $\tilde{a}\in\left[-a,a\right]$, the dynamics of the observer error can be written as
	$\dot{e}_o = \begin{bmatrix}
		-\theta_1 & 1\\
		-\theta_2+\tilde{a} & 0
	\end{bmatrix} e_o$
and it can be easily verified that Assumption~\ref{as_obs_er} holds if $\theta_1 > 0$ and $\theta_2 + \tilde{a} > 0$ for all $\tilde{a}\in\left[-a,a\right]$. 
Subsequently, we consider $a = \frac{9.81}{2}$ and $b=2$. Using the approach from \cite[Section~VI.A]{hertneck21robust_arxiv}, we have computed $n_\tpar = 23$ different parameter sets that satisfy Assumption~\ref{asum_hybrid_lyap} with $\epsilon_i \in \left[-20,0.01\right]$. Moreover, we have chosen $\theta_1 = \theta_2 = 10$. The maximum sampling interval, for which UGAS can be guaranteed for periodic sampling, is $t_{\min} = \SI{0.175}{\second}$. 

To demonstrate our approach, we compare the required number of transmissions for the proposed STC mechanism  for 1000 randomly selected initial conditions to the number of transmissions that are needed for periodic sampling with stability guarantees. 
The initial conditions for $x_p$ and $x_o$ where drawn uniformly from 
$	\begin{bmatrix}
		x_{p,1}(0,0)& x_{p,2}(0,0)& x_{o,1}(0,0)& x_{o,2}(0,0)
	\end{bmatrix}\in\left[-10,10\right]^4.$
 We consider $n_\eta = 15$ and $\eta_i(0,0) = V(x_o(0,0))$ for $i\in\left\lbrace 1,\dots,n_\eta \right\rbrace$. In the first $\SI{10}{\second}$, the proposed STC mechanism requires on average 29.8 transmissions. In comparison, for periodic sampling 56 transmissions are needed for the same time span. For trajectories with a duration of $\SI{50}{\second}$ and the same initial conditions, the proposed STC mechanism requires on average 92.4 samples whilst 282 transmissions are needed for periodic sampling, leading to reduction of the required number of transmissions by factor $\frac{1}{3}$ for the proposed STC approach. 
For the considered example, the amount of reduction is larger if a larger time span is considered. A reason for that is that the observer state $x_o$ is used to determine transmission times instead of the plant state $x_p$. It can therefore happen that the value of $V(x_o)$, i.e., of the Lyapunov function for the observer state increases significantly at the beginning as the observer state converges, leading to frequent transmissions at the beginning. An exemplary trajectory where this occurs is given in Figure~\ref{fig_states}. The respective transmission intervals are given in Figure~\ref{fig_inters}.
To compensate the initial increase of the Lyapunov function, the initial values for $\eta$ could be chosen differently at the cost of reduced convergence speed.

\begin{figure}
	\centering
		\input{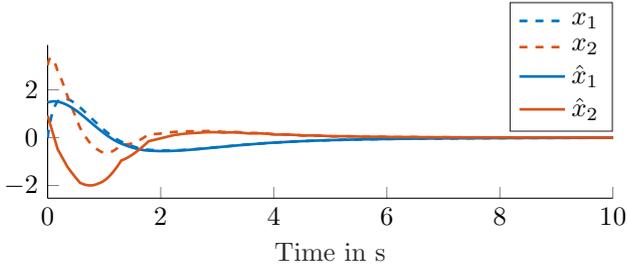}
		\vspace{-7mm}
	\caption{Exemplary trajectories of plant state and observer state for the proposed STC mechanism.
	}
	\label{fig_states}
\end{figure}
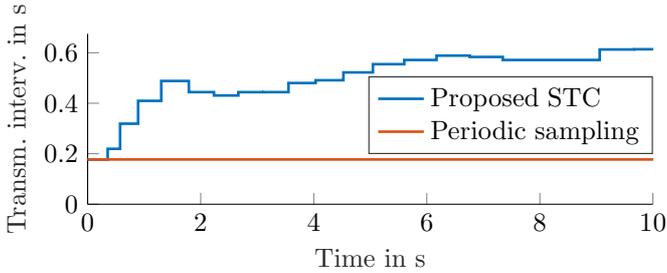
\begin{figure}
	\centering
		\definecolor{mycolor1}{rgb}{0.00000,0.44700,0.74100}%
\definecolor{mycolor2}{rgb}{0.85000,0.32500,0.09800}%
\begin{tikzpicture}

\begin{axis}[%
width=.84\linewidth,
height=.89in,
at={(0in,0in)},
scale only axis,
xmin=0,
xmax=10.0,
ymin = 0.0,
xlabel style={font=\color{white!15!black}},
xlabel={Time in \SI{}{\second}},
ylabel style={font=\color{white!15!black}},
ylabel={Transm. interv. in \SI{}{\second}},
axis background/.style={fill=white},
axis x line*=bottom,
axis y line*=left,
legend style={legend cell align=left, align=left, draw=white!15!black,at={(1,0.75)}}
]
\addplot[const plot, color=mycolor1, line width=1.0pt] table[row sep=crcr] {%
0	0.177257784781798\\
0.177257784781798	0.177257784781798\\
0.354515569563596	0.219396716821825\\
0.573912286385421	0.318966280802836\\
0.892878567188257	0.409470876897616\\
1.30234944408587	0.4881756171017\\
1.79052506118757	0.444044228823746\\
2.23456929001132	0.430741711235353\\
2.66531100124667	0.44393101538901\\
3.10924201663568	0.444044228823746\\
3.55328624545943	0.480060904901628\\
4.03334715036106	0.491071049609055\\
4.52441819997011	0.522141612224481\\
5.04655981219459	0.554735786884006\\
5.6012955990786	0.571352302216618\\
6.17264790129521	0.588622320394373\\
6.76127022168959	0.583281011304961\\
7.34455123299455	0.571352302216618\\
7.91590353521117	0.571352302216618\\
8.48725583742779	0.571352302216618\\
9.0586081396444	0.612908409606037\\
9.67151654925044	0.61381480923818\\
10.2853313584886	0.635399477042316\\
};
\addlegendentry{Proposed STC}

\addplot[const plot, color=mycolor2, line width=1.0pt] table[row sep=crcr] {%
	0	0.177257784781798\\
	11	0.177257784781798\\
};
\addlegendentry{Periodic sampling}

\end{axis}

\end{tikzpicture}%
		\vspace{-7mm}
	\caption{Time between successive transmissions for the proposed STC mechanism and periodic sampling.
	}
	\label{fig_inters}
\end{figure}
\section{Conclusion}
\label{sec_conc}
In this paper, we have presented a dynamic STC approach for output feedback in nonlinear NCS. The approach is based on hybrid Lyapunov functions and a dynamic variable that encodes past state information to determine sampling instants. Instead of the plant state, the state of an asymptotically stable continuous-time observer is used to determine transmission times. 
The design of the STC mechanism is decoupled from the observer design and the same assumptions on the plant as for the full state-feedback case are used. Stability guarantees were derived and the proposed approach was illustrated with
 a numerical example. 
\bibliography{../../../../Literatur/literature}
\arxiv{}{\appendix
\section{}
\label{app_a}
	\textbf{Proof of Theorem~\ref{prop_fir}}. Consider any maximal solution $\xi$ to \eqref{eq_sys_hyb}. 
	Note that $\hat{x}_o(\tvar_j^+) = x_o(\tvar_j^+) = x_o(\tvar_j)$ and thus $e(\tvar_j^+) = e_o(\tvar_j^+) = e_o(\tvar_j)$.
	Obviously, $\bar{h} \geq \delta T_{\max}(\gamma_1,L_1+\frac{\epsilon_1}{2}) = t_{\min}$ in Algorithm~\ref{algo_trig_window} and thus $\svar_{j+1}-\svar_j = \Gamma(x_o(\tvar_j),\eta(\tvar_j)) \geq t_{\min}$ holds for all $j\in\dom~\xi$, i.e. the minimum time between two triggering instants is strictly positive. 
	Because of the update of $\bar{h}$ in Algorithm~\ref{algo_trig_window}, for each $j\in\dom~\xi$, there is  an $\ivar \in\left\lbrace 1,\dots, n_\tpar \right\rbrace$ such that  $\svar_{j+1}-\svar_j \leq T_{\max}\left(\gamma_\ivar,\max\left\lbrace L_\ivar+\frac{\epsilon_\ivar}{2}, 1-\delta \right\rbrace\right)$ and $\svar_{j+1}-\svar_j \leq t_{\max} \coloneqq \underset{i\in\left\lbrace 1,\dots,n_{\tpar}\right\rbrace}{\max} \delta T_{\max}(\gamma_i,\max\left\lbrace L_i+\frac{\epsilon_i}{2}, 1-\delta \right\rbrace)$. 
	Proposition~\ref{prop_hybrid} thus implies for $\svar_j\leq t \leq \svar_{j+1}$ with $x_p(\tvar_{j}^+) = x_p(\tvar_{j})$ and $e(\tvar_j^+) = e_o(\tvar_j)$ that 
	\begin{equation}
		\label{eq_V_dec_fir}
		\begin{split}
			&V(x_p(\svar,j)) \leq U_\ivar(\xi(t,j\change{+1}{}))	
			\leq 	e^{ -\epsilon_\ivar (t-\svar_j)}U_\ivar(\xi(\tvar_j^+))\\ %
			=& e^{ -\epsilon_\ivar (t-\svar_j)}\left(V(x_p(\tvar_j)) + \gamma_\ivar \lambda_\ivar^{-1} W^2(e_o(\tvar_j)) \right).	
		\end{split}
	\end{equation}
	Here $U_\ivar(\xi)$ is the respective function according to \eqref{eq_def_u} from Proposition~\ref{prop_hybrid} for $\gamma = \gamma_\ivar$, $L = L_\ivar, \epsilon = \epsilon_\ivar$ and some sufficiently small $\lambda\in\left(0,1\right)$.

	We will now use this to investigate the evolution of $V(x_p)$ depending on the time between sampling instants. We distinguish between two possible outcomes for $\ivar$ in Algorithm~\ref{algo_trig_window}.
	Suppose first $\ivar = 1$, i.e., the fallback strategy is used. Then
	\begin{equation*}
		e^{-\epsilon_\ivar(\svar_{j+1}-\svar_{j})} V(x_p(\tvar_j)) \leq e^{-\epsilon_\refer(\svar_{j+1}-\svar_{j})} V(x_p(\tvar_j)) 
	\end{equation*}
	holds since $\epsilon_1 \geq \epsilon_\refer > 0$ and $\svar_{j+1}-\svar_j < T_{\max}(\gamma_1,L_1+\frac{\epsilon_1}{2})$. This implies with \eqref{eq_V_dec_fir} that 
	\begin{equation}
		\label{eq_bound_i1}
		\begin{split}
			V(x_p(\tvar_{j+1})) 
			\leq& e^{-\epsilon_\refer(\svar_{j+1}-\svar_{j})} V(x_p(\tvar_j))  + \alpha_1(\abs{e_o(\tvar_j)})	
		\end{split}
	\end{equation}
	for some $\alpha_1\in\mathcal{K}$. Here we used that $\underset{i\in\left\lbrace 1,\dots,n_{\tpar}\right\rbrace}{\max}e^{ -\epsilon_i (\svar_{j+1}-\svar_j)} \in \R_{>0}$ due to the upper bound $t_{\max}$ on $t_{j+1}-t_j$. 
		
	Next, suppose that $\ivar > 1$. In this case, Algorithm~\ref{algo_trig_window} chooses $\svar_{j+1}-\svar_j$ such that $\svar_{j+1}-\svar_j < T_{\max}(\gamma_\ivar,L_\ivar+\frac{\epsilon_\ivar}{2})$ and
	\begin{equation}
		\label{eq_Vo_bound}
		\begin{split}
			e^{-\epsilon_\ivar(\svar_{j+1}-\svar_{j})} V(x_o(\tvar_j)) \leq& e^{-\epsilon_\refer(\svar_{j+1}-\svar_j)} C(x_o(\tvar_j),\eta(\tvar_j))\\
		\end{split}		
	\end{equation}
	hold. Further note that $\ivar > 1$ is only possible if $V(x_o(\tvar_j)) \leq C(x_o(\tvar_j),\eta(\tvar_j)) \leq V_{\max}$ due to Line~\ref{line_for_start} of Algorithm~\ref{algo_trig_window} and our choice of $C(x_p,\eta)$ according to \eqref{eq_C_def}.
Observe that
	\begin{equation*}
		\begin{split}
			V(x_p(\tvar_j)) =& V(x_o(\tvar_j) - e_o(\tvar_j))\\
			\leq& V(x_o(\tvar_j)) +\mathfrak{L}(\abs{e_o(\tvar_j)}) \abs{e_o(\tvar_j)}
		\end{split}
	\end{equation*}
	where $\mathfrak{L}(p)$ is a (local) Lipschitz constant of $V$ that satisfies
	\begin{equation}
		\label{eq_def_L}
	\abs{V(x_o(\tvar_j)-e_o(\tvar_j)) - V(x_o(\tvar_j))}\leq \mathfrak{L}(p) \abs{e_o(\tvar_j)}
	\end{equation}
	for all $e_o(\tvar_j)$ with $\abs{e_o(\tvar_j)} \leq p$ and $x_o$ with $V(x_o) \leq V_{\max}$. Note that $\mathfrak{L}(p)$ is non-decreasing and bounded for all $p \geq 0$ since $V$ is locally Lipschitz. We thus obtain for some  $\alpha_{2}\in\mathcal{K}$
	\begin{equation}
		\label{eq_bound_op}
		V(x_p(\tvar_j)) \leq V(x_o(\tvar_j)) + \alpha_{2}(\abs{e_o(\tvar_j)}).
	\end{equation}
	Using \eqref{eq_Vo_bound} and \eqref{eq_bound_op} we obtain from \eqref{eq_V_dec_fir} for $t = t_{j+1}$
	\begin{equation}
		\label{eq_bound_ij}
		\begin{split}
			&V(x_p(\tvar_{j+1}))	\\
			\leq& e ^{-\epsilon_\refer(t_{j+1}-t_j)} C(x_o(\tvar_j),\eta(\tvar_j))\\
			 &+ e^{ -\epsilon_\ivar (\svar_{j+1}-\svar_j)}\left( \alpha_{2}(\abs{e_o(\tvar_{j})}) + \gamma_\ivar \lambda_\ivar^{-1} W^2(e_o(\tvar_j)) \right)	\\
			\leq& e ^{-\epsilon_\refer(t_{j+1}-t_j)} C(x_o(\tvar_j),\eta(\tvar_j)) + \alpha_3(\abs{e_o(\tvar_j)})\\
		\end{split}
	\end{equation}
for some $\alpha_3\in\mathcal{K}$, where we used again that \linebreak $\underset{i\in\left\lbrace 1,\dots,n_{\tpar}\right\rbrace}{\max}e^{ -\epsilon_i (\svar_{j+1}-\svar_j)} \in \R_{>0}$ due to the upper bound $t_{\max}$ on $t_{j+1}-t_j$. 
From \eqref{eq_C_def}, we obtain with the update of $\eta$ according to \eqref{eq_S_window_2} that\footnote{Note that the second sum is only relevant for $j < m-1$ to capture the effect of the initial condition on $\eta$.}
\begin{equation}
	\label{eq_C_decomp}
	\begin{split}
		&C(x_o(\tvar_{{j}}),\eta(\tvar_{{j}})) = \frac{1}{m} V(x_o(\tvar_{{j}})) + \sum_{k=1}^{m-1} \eta_k(\tvar_{{j}}) \\
		\leq & \frac{1}{m}V(x_o(\tvar_{{j}}))\\
		 &+ \frac{1}{m}\sum_{k=1}^{\min\left\lbrace m-1 , j \right\rbrace} e^{-\epsilon_\refer(t_{{j}}-t_{{j}-k})} \min\left\lbrace V(x_o(\tvar_{{{j}}-k})), V_{\max} \right\rbrace\\
		 &+ \frac{1}{m}\sum_{k = \min\left\lbrace j,m-1\right\rbrace+1}^{m-1} e^{-\epsilon_\refer(t_{{j}}-t_{0})}\eta_{m-k}.
	\end{split}
\end{equation}

If $V(x_o(\tvar_{j-k})) \leq V_{\max},$ we obtain similar as in \eqref{eq_bound_op} that
\begin{equation*}
	\begin{split}
			&\min\left\lbrace V(x_o(\tvar_k)), V_{\max} \right\rbrace\\
			 =& V(x_p(\tvar_{j-k})) + V(x_o(\tvar_{j-k})) - V(x_o(\tvar_{j-k}) - e_o(\tvar_{j-k}))\\
		\leq& V(x_p(\tvar_{j-k})) + \alpha_{2}(\abs{e_o(\tvar_{j-k})}).
	\end{split}
\end{equation*} 
If $V(x_o(\tvar_{j-k})) > V_{\max},$ then either $V(x_p(\tvar_{j-k})) > V_{\max}$ and $\min\left\lbrace V(x_o(\tvar_{j-k})), V_{\max} \right\rbrace \leq V(x_p(\tvar_{j-k})) + \alpha_{2}(\abs{e_o(\tvar_{j-k})})$ trivially holds or $V(x_p(\tvar_{j-k})) \leq V_{\max}$. In the latter case, we can again use the same argumentation that precedes \eqref{eq_bound_op} and obtain
\begin{equation}
	\label{eq_bound_op2}
	\min\left\lbrace V(x_o(\tvar_{j-k})), V_{\max} \right\rbrace \leq V(x_p(\tvar_{j-k})) + \alpha_{2}(\abs{e_o(\tvar_{j-k})}).
\end{equation}
Thus \eqref{eq_bound_op2} holds in all cases.
Using it in \eqref{eq_C_decomp}, we obtain 
\begin{equation}
	\label{eq_C_bound_pre}
	\begin{split}
		&C(x_o(\tvar_{{j}}),\eta(\tvar_{{j}}))\\
		 \leq& \frac{1}{m}\sum_{k=1}^{\min\left\lbrace m-1 , j \right\rbrace} e^{-\epsilon_\refer(t_{{j}}-t_{{j}-k})} V(x_p(\tvar_{{{j}}-k}))\\
		  &+\frac{1}{m}\sum_{k=1}^{\min\left\lbrace m-1 , j \right\rbrace} e^{-\epsilon_\refer(t_{{j}}-t_{{j}-k})}  \alpha_{2}(\abs{e_o(\tvar_{j-k})})\\
		  &+ \frac{1}{m}\sum_{k = \min\left\lbrace j,m-1\right\rbrace+1}^{m-1} e^{-\epsilon_\refer(t_{{j}}-t_{0})}\eta_{m-k}.
	\end{split}
\end{equation}		  
With \eqref{eq_obs_er} from Assumption~\ref{as_obs_er} in the second sum of \eqref{eq_C_bound_pre}, and $t_{k+1}-t_k \leq t_{\min}~\forall k\in\dom\xi$, we further obtain
\begin{equation}
	\label{eq_C_bound}
	\begin{split}		  
		 &C(x_o(\tvar_{{j}}),\eta(\tvar_{{j}}))\\
		   \leq & \frac{1}{m}\sum_{k=1}^{\min\left\lbrace m-1 , j \right\rbrace} e^{-\epsilon_\refer(t_{{j}}-t_{{j}-k})} V(x_p(\tvar_{{{j}}-k}))\\
		   	   &+ \frac{1}{m}\sum_{k = \min\left\lbrace j,m-1\right\rbrace+1}^{m-1} e^{-\epsilon_\refer(t_{{j}}-t_{0})}\eta_{m-k}\\
		   	   &+    \alpha_{2}\left(\beta_o\left(\abs{e_o(\tnn)},\max\left\lbrace (j-m+1)t_{\min},0\right\rbrace\right)\right).
	\end{split}
\end{equation}

	Now we show by induction based on \eqref{eq_bound_i1} and \eqref{eq_bound_ij} that
	\begin{equation}
		\label{eq_ind_as}
		\begin{split}
			&V(x_p(\tvar_j)) 
			\\ \leq& e^{-\epsilon_\refer t_j} \max\left\lbrace V(x_p(0,0)),\abs{\eta(0,0)}\right\rbrace\\
			&+ \sum_{k=0}^{j-1}\left(e^{-\epsilon_\refer t_{\min}(j- k -1)}\right. \\
			&\cdot \left.\alpha_5\left(\beta_o\left(\abs{e_o(\tnn)},\max\left\lbrace (k-m+1)t_{\min},0\right\rbrace\right)\right) \vphantom{e^{-\epsilon_\refer t_{\min}(j- k)}}\right)
		\end{split}		
	\end{equation}
	holds for all $j\in\dom~\xi$ with $\alpha_5(\cdot) \coloneqq \alpha_1(\cdot)+\alpha_{2}(\cdot)+\alpha_3(\cdot) \in\mathcal{K}$. It trivially holds for $j = 0$. Suppose it holds for all $j \leq \tilde{j}$ for some $\tilde{j}\in\dom~\xi$. 
	
	We first consider the case $i_{\tilde{j}} \geq 1$. Note that \eqref{eq_ind_as} for $j\leq \tilde{j}$ implies with \eqref{eq_C_bound} that
	\begin{equation*}
		\begin{split}
			&C(x_o(\tvar_{\tilde{j}}),\eta(\tvar_{\tilde{j}}))\\
			\leq&  e^{-\epsilon_\refer t_{\tilde{j}}} \max\left\lbrace V(x_p(0,0)),\abs{\eta(0,0)}\right\rbrace\\
			&+ \sum_{k=0}^{{\tilde{j}-1}}\left(e^{-\epsilon_\refer t_{\min}({\tilde{j}}- k-1)}\right. \\
			&~\cdot \left.\alpha_5\left(\beta_o\left(\abs{e_o(\tnn)},\max\left\lbrace (k-m+1)t_{\min},0\right\rbrace\right)\right) \vphantom{e^{-\epsilon_\refer t_{\min}(j- k)}}\right)\\
			&+ \alpha_{2}\left(\beta_o\left(\abs{e_o(\tnn)},\max\left\lbrace (\tilde{j}-m+1)t_{\min},0\right\rbrace\right)\right).
		\end{split}
	\end{equation*}
	Plugging this in \eqref{eq_bound_ij}, we obtain
	\begin{equation}
		\label{eq_ind_ii}
		\begin{split}
			&V(x_p(\tvar_{\tilde{j}+1})) \\
			\leq& e^{-\epsilon_\refer (t_{\tilde{j}+1}-t_{\tilde{j}})} e^{-\epsilon_\refer t_{\tilde{j}}} \max\left\lbrace V(x_p(0,0)),\abs{\eta(0,0)}\right\rbrace\\
			& + \alpha_3\left(\abs{e_o(\tvar_{\tilde{j}})}\right)
			+ e^{-\epsilon_\refer (t_{\tilde{j}+1}-t_{\tilde{j}})} \sum_{k=0}^{{\tilde{j}-1}}\left(e^{-\epsilon_\refer t_{\min}({\tilde{j}}- k-1)}\right. \\
			&~\cdot \left.\alpha_5\left(\beta_o\left(\abs{e_o(\tnn)},\max\left\lbrace (k-m+1)t_{\min},0\right\rbrace\right)\right) \vphantom{e^{-\epsilon_\refer t_{\min}(j- k)}}\right)\\
			&+  e^{-\epsilon_\refer (t_{\tilde{j}+1}-t_{\tilde{j}})}\alpha_{2}\left(\beta_o\left(\abs{e_o(\tnn)},\max\left\lbrace (\tilde{j}-m+1)t_{\min},0\right\rbrace\right)\right).
		\end{split}
	\end{equation}
	Note that
	\begin{equation*}
		\begin{split}
			& \alpha_3\left(\abs{e_o(\tvar_{\tilde{j}})}\right)\\
			 &+ e^{-\epsilon_\refer (t_{\tilde{j}+1}-t_{\tilde{j}})}\alpha_{2}\left(\beta_o\left(\abs{e_o(\tnn)},\max\left\lbrace (\tilde{j}-m+1)t_{\min},0\right\rbrace\right)\right)\\
			\leq & \alpha_5(\beta_o\left(\abs{e_o(\tnn)},\max\left\lbrace (\tilde{j}-m+1)t_{\min},0\right\rbrace\right))
		\end{split}
	\end{equation*}
holds since 
\begin{equation*}
	\begin{split}
		\abs{e_o(\tvar_{\tilde{j}})} &\leq \beta_o\left(\abs{e_o(\tnn)},\svar_{\tilde{j}}\right)\\
		 &\leq \beta_o\left(\abs{e_o(\tnn)},\max\left\lbrace (\tilde{j}-m+1)t_{\min},0\right\rbrace\right),
	\end{split}
\end{equation*} since $e^{-\epsilon_\refer (t_{\tilde{j}+1}-t_{\tilde{j}})} < 1$ and due to the definition of $\alpha_5$. 
	
	Using this and  $e^{-\epsilon_\refer(\svar_{\tilde{j}+1}-\svar_{\tilde{j}})} \leq e^{-\epsilon_\refer t_{\min}} $ in \eqref{eq_ind_ii}, we obtain 
	\begin{equation*}
		\begin{split}
			&V(x_p(\tvar_{\tilde{j}+1})) \\
			\leq & e^{-\epsilon_\refer t_{\tilde{j}+1}} \max\left\lbrace V(x_p(0,0)),\abs{\eta(0,0)}\right\rbrace\\
			&+ \sum_{k=0}^{{\tilde{j}}}\left(e^{-\epsilon_\refer t_{\min}({\tilde{j}}- k)}\right. \\
			&~\cdot \left.\alpha_5\left(\beta_o\left(\abs{e_o(\tnn)},\max\left\lbrace (k-m+1)t_{\min},0\right\rbrace\right)\right) \vphantom{e^{-\epsilon_\refer t_{\min}(j- k)}}\right),
		\end{split}
	\end{equation*}
	i.e., \eqref{eq_ind_as} also holds for $j=\tilde{j}+1$ in this case.

	Now we consider the remaining case $i_{\tilde{j}} = 1$. In this case, \eqref{eq_bound_i1} holds and hence with \eqref{eq_ind_as} for $j=\tilde{j}$, we obtain
	\begin{equation*}
		\begin{split}
			&V(x(\tvar_{\tilde{j}+1}))\\ \leq& e^{-\epsilon_\refer(\svar_{\tilde{j}+1}-\svar_{\tilde{j}})} V(x_p(\tvar_{\tilde{j}}))  + \alpha_1\left(\abs{e_o(\tvar_{\tilde{j}})}\right)	\\
			\leq& e^{-\epsilon_\refer t_{\tilde{j}+1}} \max\left\lbrace V(x_p(0,0)),\abs{\eta(0,0)}\right\rbrace\\
			&+ \sum_{k=0}^{\tilde{j}}\left(e^{-\epsilon_\refer t_{\min}(\tilde{j}- k)}\right. \\
			&\cdot \left.\alpha_5\left(\beta_o\left(\abs{e_o(\tnn)},\max\left\lbrace (k-m+1)t_{\min},0\right\rbrace\right)\right) \vphantom{e^{-\epsilon_\refer t_{\min}(j- k)}}\right)
		\end{split}
	\end{equation*}
	where we used that $ e^{-\epsilon_\refer t_{\tilde{j}+1}} \leq e^{-\epsilon_\refer t_{\min}}$ and that
	\begin{equation*}
		\begin{split}
			&\alpha_1\left(\abs{e(\tvar_{\tilde{j}})}\right) \leq \alpha_1\left(\beta_o\left(\abs{e_o(\tnn)},\svar_{\tilde{j}}\right)\right)  \\
			 \leq& \alpha_1\left(\beta_o\left(\abs{e_o(\tnn)},\max\left\lbrace (\tilde{j}-m+1)t_{\min},0\right\rbrace\right)\right)\\
			 \leq&  \alpha_5\left(\beta_o\left(\abs{e_o(\tnn)},\max\left\lbrace (\tilde{j}-m+1)t_{\min},0\right\rbrace\right)\right).
		\end{split}
	\end{equation*}
	 As a result, we can conclude that $\eqref{eq_ind_as}$ holds also for $j = \tilde{j}+1$ if $i_{\tilde{j}} = 1$.
	
	It thus follows by induction that \eqref{eq_ind_as} holds for all $j\in\dom~\xi$. Together with the fact that $t_{\min}\leq t_{j+1}-t_j\leq t_{\max}$ this further implies that $\xi$ is $t$-complete.  
	
	Next, we discuss that \eqref{eq_ind_as} implies that $U_{\ivar}$ is bounded by a $\mathcal{K}\mathcal{L}\mathcal{L}$ function. 
	Observe 
	\newlength\mylen
	\settoheight\mylen{$+ \frac{1}{1-e^{-\epsilon_\refer t_{\min}}} \beta_o\left(\abs{e_o(\tnn)},\max\left\lbrace \left(\frac{t}{2t_{\max}}-m+1\right)t_{\min},0\right\rbrace\right)$}
	\begin{equation}
		\label{eq_sum_bound_1}
		\begin{split}
			&\sum_{k=0}^{j-1}e^{-\epsilon_\refer t_{\min}(j-k-1)}\\
			&\cdot \alpha_5\left(\beta_o\left(\abs{e_o(\tnn)},\max\left\lbrace (k-m+1)t_{\min},0\right\rbrace\right)\right) \\			
		\leq&  e^{-\epsilon_\refer t_{\min} \frac{j}{2}} \alpha_5({\beta_o(\abs{e(\tnn)},0)}) \sum_{k=0}^{\frac{j}{2}}e^{-\epsilon_\refer t_{\min}({\frac{j}{2}}- k-1)}\\
		&+ \alpha_5\left(\beta_o\left(\abs{e(\tnn)},\max\left\lbrace \left(\frac{j}{2}-m+1\right)t_{\min},0\right\rbrace\right)\right)\\
		&\cdot \sum_{k=\frac{j}{2}}^{j-1}e^{-\epsilon_\refer t_{\min}({j}- k-1)}\\
		\leq& e^{-\epsilon_\refer \frac{t_{\min}}{t_{\max}} \frac{t}{2}} \frac{1}{1-\epsilon_\refer t_{\min}}\beta_o(\abs{e_o(\tnn)},0)\\
		&\resizebox{\linewidth}{.95\mylen}{$+ \frac{1}{1-e^{-\epsilon_\refer t_{\min}}} \beta_o\left(\abs{e_o(\tnn)},\max\left\lbrace \left(\frac{t}{2t_{\max}}-m+1\right)t_{\min},0\right\rbrace\right)$}\\
		\eqqcolon& \beta_2(\abs{e_o(\tnn)},t) \in \mathcal{K}\mathcal{L},
		\end{split}
	\end{equation} 
	where we used the geometric series and the fact that $j \geq \frac{t}{t_{\max}}$ since $t_{j+1}-t_j \leq t_{\max}$. 
	
	Plugging this into \eqref{eq_ind_as}, we can conclude that for some $\beta_3\in\mathcal{K}\mathcal{L}$,
	$	V(x(\tvar_j)) \leq \beta_3\left(\abs{
			\begin{bmatrix} 				
				x_p(0,0)\\ 
				e(0,0)\\ 				
				\eta(0,0) 		\end{bmatrix}},\svar_j\right)$
	holds for all $j\in\R_{>0}$. Using again \eqref{eq_V_dec_fir}, this implies together with Assumption~\ref{as_obs_er} for $k_1 = \underset{i\in\left\lbrace 1,\dots,n_{\tpar}\right\rbrace}{\max}e^{ -\epsilon_i (\svar_{j+1}-\svar_j)} \in \R_{>0}$ for some $\beta_4 \in\mathcal{K}\mathcal{L}\mathcal{L}$ that
	\begin{equation*}
		\begin{split}
			U_\ivar(t,j) \leq& k_1 \left(\beta_3\left(\abs{
				\begin{bmatrix} 				
					x_p(0,0)\\ 
					e(0,0)\\ 				
					\eta(0,0) 		\end{bmatrix}},\max\left\lbrace t - t_{\max}, 0 \right\rbrace\right)\right.\\
				&\left.+ \gamma_\ivar\lambda_\ivar^{-1} W^2(\beta_0(\abs{e_0(0,0)}, \max\left\lbrace t - t_{\max}, 0 \right\rbrace)  \vphantom{\begin{bmatrix} 				
						x_p(0,0)\\ 
						e(0,0)\\ 				
						\eta(0,0) 		\end{bmatrix}}\right)\\
					&\leq \beta_4\left(\begin{bmatrix} 				
						x_p(0,0)\\ 
						e(0,0)\\ 				
						\eta(0,0) 		\end{bmatrix},t,j\right),
		\end{split}		
	\end{equation*}
	where we used that $t \geq \frac{t}{2} + jt_{\min}$ holds for all $(t,j)\in\dom~\xi$. 
	
	Finally, with the bounds on $V$ and $W$ from Assumption~\ref{asum_hybrid_lyap}, the fact that $\phi_\ivar \in \left[\lambda,\lambda^{-1}\right]$ for sufficient small $\lambda\in\left(0,1\right)$, the definition of $U_\ivar$ according to \eqref{eq_def_u} for the respective $\gamma_\ivar$, $L_\ivar$ and $\epsilon_\ivar$ and since the observer state converges to the plant state according to Assumption~\ref{as_obs_er}, UGAS of the set $\left\lbrace \left(x_p,x_c,e,\eta,\tau,\auxvar\right): x_p = 0, x_c = 0, e= 0, \eta = 0 \right\rbrace$ follows similar as in \cite{hertneck21robust_arxiv}. \hfill\hfill\qed

}

\end{document}